\documentclass[12pt]{iopart}

\usepackage{graphicx,setspace}
\usepackage{dsfont}
\usepackage{iopams}
\usepackage{mathrsfs}
\usepackage{psfrag}
\usepackage{cases}

\begin{document}

\title{Jarzynski equality for a gas particle driven out of equilibrium}
\author{T G Philbin and J Anders}
\address{Department of Physics and Astronomy, University of Exeter,
Stocker Road, Exeter EX4 4QL, UK.}
\eads{\mailto{t.g.philbin@exeter.ac.uk} }

\begin{abstract}
One particle in a classical perfect gas is driven out of equilibrium by changing its mass over a short time interval. The work done on the driven particle depends on its collisions with the other particles in the gas. This model thus provides an example of a non-equilibrium process in a system (the driven particle) coupled to an environment (the rest of the gas). We calculate the work done on the driven particle and compare the results to Jarzynski's equality relating a non-equilibrium work process to an equilibrium free-energy difference. The results for this model are generalised to the case of a system that is driven in one degree of freedom while interacting with the environment through other degrees of freedom. 
\end{abstract}


\section{Introduction}
Relations between equilibrium thermodynamic quantities and the work done in non-equilibrium processes are standard textbook material~\cite{LLsp1}. For example,  in a non-equilibrium process the average work done $\langle w \rangle$ on a thermal ensemble of systems with a fixed volume obeys (\cite{LLsp1}, \S~20) 
\begin{equation}  \label{wDF}
\langle w \rangle \geq \Delta F , 
\end{equation}
where $\Delta F$ is the change in the Helmholtz free energy of the system after the final state equilibrates at the same temperature as the initial state. This inequality follows from the second law of thermodynamics~\cite{LLsp1}. The process by which work is done on the system corresponds to a time variation of some of its state parameters and this variation must be identical for each ensemble element. Because of the meaning of $\Delta F$ in (\ref{wDF}), one often envisages the experimental system re-equilibrating after the work is done, so that the final state, as well as the initial state, is in thermal equilibrium. This assumption of a final equilibrium state is not necessary in order to use (\ref{wDF}), however: the work can be performed and measured without letting the system re-equilibrate and (\ref{wDF}) still gives a prediction regarding the average measured work that can be verified if one can calculate the equilibrium free-energy difference. 

In recent times fluctuation theorems have generalised this inequality to a set of non-equilibrium \emph{equalitites}. Detailed fluctuation relations show that the probability distributions of stochastically fluctuating quantities for a non-equilibrium process, such as entropy, work and heat, are linked to equilibrium properties and corresponding quantities for the time-reversed process~\cite{Evans93, Crooks99, Kawai07}. A well-known integral fluctuation relation is the Jarzynski work relation where the exponentiated work is averaged over its distribution and related to an equilibrium free-energy difference~\cite{jar97,jar04}. Jarzynski's relation strengthens (\ref{wDF}) by including all moments of the non-equilibrium work resulting in an \emph{equality} from which (\ref{wDF}) follows for the first moment.  Experiments with bio-molecules have used fluctuation theorems to derive, from the measurable work in non-equilibrium pulling experiments, the desired equilibrium free-energy surface of the molecules~\cite{Liphardt02, Collin05}, the latter being impossible to measure directly. In physics, fluctuation theorems have been measured for example, for a defect center in diamond~\cite{Schuler05}, for a torsion pendulum~\cite{Douarche06}, and in an electronic system~\cite{Saira12}. Extensions to the quantum regime are reviewed in~\cite{Esposito09} and the first quantum experiments using nuclear magnetic resonance are reported in~\cite{Batalhao13}.

Jarzynski's relation in its general form states
\begin{equation}  \label{jar}
\left\langle \exp\left(-\frac{w}{k_BT}\right) \right\rangle = \exp\left(-\frac{\Delta F^\star}{k_BT}\right) ,
\end{equation}
where $T$ is the temperature of the initial thermal ensemble and $\Delta F^\star$ denotes the change in a free-energy measure $F^\star$ that differs from the usual Helmholtz free energy by the inclusion of a contribution from the coupling of the system to the thermal environment (heat reservoir)~\cite{jar04}. The final value of $F^\star$ used to calculate $\Delta F^\star$ is that of the final state of the system after it equilibrates at the same temperature ($T$) as the initial state. The usual assumption of quasi-closed systems~\cite{LLsp1} is based on a system-environment coupling that gives a very small contribution to the system Hamiltonian leading to thermodynamic quantities that are extensive~\cite{abe}. Quantities such as the Helmholtz free energy $F$ are then calculated using the free-system Hamiltonian, without taking account of the coupling to the environment~\cite{LLsp1}. This is often an excellent approximation, but even then the Jarzynski equality, as a strict equality, does not hold for the Helmholtz free-energy difference $\Delta F$, but rather for $\Delta F^\star$. For significant system-environment coupling there will in general be no close relation between $\Delta F^\star$ and the free-energy difference $\Delta F$ calculated from the Hamiltonian of the uncoupled system. 

The free energy $F^\star$ is defined as follows~\cite{jar04}. Let the total Hamiltonian of the system and environment be
\begin{equation}  \label{H}
H(x,X)=H_s(x)+H_e(X)+H_\mathrm{int}(x,X),
\end{equation}
where $H_s(x)$ is the Hamiltonian of the free system, whose canonical variables are $x=\{q_i,p_i\}$, $H_e(X)$ is the Hamiltonian of the free environment, whose canonical variables are $X=\{Q_i,P_i\}$, and $H_\mathrm{int}(x,X)$ is the system-environment interaction term. If the coupled system and environment are in thermal equilibrium at temperature $T$ then it is easy to show that the system is distributed according to a phase space probability density $\rho(x)$ that can be written as
\begin{equation}   \label{rho}
\rho(x)=\frac{1}{Z^\star}\exp\left(-\frac{H^\star(x)}{k_BT}\right) ,
\end{equation}
where $H^\star(x)$ is an effective Hamiltonian for the system~\cite{Gelin09} known as the Hamiltonian of mean force~\cite{kir35, Campisi09}. The definition of $H^\star(x)$ is
\begin{equation}  \label{Hmf}
\fl
H^\star(x)=H_s(x)-\frac{1}{k_BT}\ln\left[\frac{\int dX\,\exp\left\{-\left[H_e(X)+H_\mathrm{int}(x,X)\right]/(k_BT)\right\}}{\int dX\,\exp\left[-H_e(X)/(k_BT)\right]}\right]
\end{equation}
and $Z^\star$ is the partition function associated with the distribution (\ref{rho}):
\begin{equation}  \label{Zstar}
Z^\star=\int dx\,\exp\left(-\frac{H^\star(x)}{k_BT}\right).
\end{equation}
The distribution (\ref{rho}) of the system is obtained by ``tracing out" the environment; the coupling of the system to the environment means this distribution is not a Boltzmann distribution based on the free-system Hamiltonian $H_s(x)$. The choice of $H^\star(x)$ as the effective Hamiltonian in the distribution (\ref{rho}) is not unique: one could add any $x$-independent term to (\ref{Hmf}) (for example one could cancel the denominator in the logarithm) and re-define the partition function (\ref{Zstar}) accordingly. But $H^\star(x)$ reduces to the free-system Hamiltonian $H_s(x)$ when the coupling to the environment vanishes, which is a partial justification for its use as an effective Hamiltonian of the system.\footnote{There are other reasons to view $H^\star(x)$ as the effective Hamiltonian of the system~\cite{jar04}. A perhaps surprising validation of the Hamiltonian of mean force occurs in the Casimir effect, where it determines the zero-point and thermal energy density of electromagnetic fields inside materials~\cite{phi11}. Casimir forces calculated from this energy density agree with those deduced from the electromagnetic stress tensor~\cite{phi11}.} The free energy $F^\star$ that appears in the Jarzynski equality (\ref{jar}) is the free energy associated with the partition function (\ref{Zstar}):
\begin{equation}  \label{Fstar}
F^\star=-k_BT\ln Z^\star.
\end{equation}

The purpose of this paper is to demonstrate the Jarzynski equality in a model system, an exercise that is nontrivial except in the least interesting case of an isolated system. As Jarzynski remarks in~\cite{jar04}, ``Exactly solvable models are hard to come by!", and we aim to provide one. Our choice of model is the perfect gas, one of the classic thermodynamic systems. A single particle in the classical perfect gas is treated in the textbooks as a quasi-closed system, so that the thermodynamic quantities for the particle, and hence for the entire gas, are calculated using the free-particle Hamiltonian~\cite{LLsp1}. In reality however, the single particle interacts with the other particles through collisions; all the other particles act as a thermal environment to which the single particle is coupled. Our interest in this model arises from the fact that the system we will consider (a single particle) is coupled to the environment (the other particles) whereas the Helmholtz free energy of the system ignores this coupling. The question then arises of the difference between the Helmholtz free energy $F$ and the free-energy $F^\star$ that appears in Jarzynski's equality (\ref{jar}) for systems coupled to the environment. Anticipating our results, we will find that although $F^\star\neq F$ because of the system-environment coupling,  $\Delta F^\star$ will be equal to  $\Delta F$ for the particular non-equilibrium process that we will analyse. The model reveals an interesting general implication from its structure: the Jarzynski equality holds in the form
\begin{equation}  \label{jar2}
\left\langle \exp\left(-\frac{w}{k_BT}\right) \right\rangle = \exp\left(-\frac{\Delta F}{k_BT}\right) ,
\end{equation}
where $F$ is the Helmholtz free energy calculated from the free-system Hamiltonian, in the general case of a system that is driven in one degree of freedom while interacting with the environment through other degrees of freedom. This is valid \emph{even when the coupling to the environment is arbitrarily large}.

\section{Model}

In the case of a perfect gas the total Hamiltonian (\ref{H}) of the system (a single particle) plus the environment (all the other particles) is given by
\begin{equation}  \label{Hgas}
\fl
H_s(\bi{p})=\frac{p^2}{2m}, \qquad H_e(\{\bi{P}_i\})=\sum_i\frac{P_i^2}{2m}, \qquad H_\mathrm{int}(\bi{q},\{\bi{Q}_i\})=V(\bi{q},\{\bi{Q}_i\}).
\end{equation}
The free-system and free-environment Hamiltonians ($H_s$ and $H_e$, respectively) depend on the canonical momenta of the particles while the interaction potential depends on the particle positions. We consider a gas of finite volume so the interaction potential $V(\bi{q},\{\bi{Q}_i\})$ describes not just the collisions between the particles but also the collisions with the bounding walls (the walls are thus also part of the environment to which the system is coupled). Collisions between the particles in our classical model will have to be calculated exactly so we take them to be impenetrable spheres of radius $R$. The interaction potential $V(\bi{q},\{\bi{Q}_i\})$ thus consists of a set of potential barriers created by the particles and the walls; the former barriers are functions of the distances between pairs of particles, while the latter barriers depend on the distances between the particles and the walls.

The gas is initially in thermal equilibrium at temperature $T$. In the usual approximation where each particle is considered as a quasi-closed system~\cite{LLsp1}, the coupling between the particles is ignored and the particle forming our system of interest occupies an infinitesimal volume of its phase space centred on $(\bi{q},\bi{p})$ with a probability 
\begin{equation}  \label{bolzatom}
\fl
\rho(\bi{q},\bi{p}) \,\rmd^3q\,\rmd^3p=Z^{-1}\exp\left(-\frac{p^2}{2mk_BT}\right)\,\rmd^3q\,\rmd^3p, \qquad Z=V\left(2\pi mk_BT\right)^{3/2},
\end{equation}
where $Z$ is the partition function and $V$ is the volume of the gas. The probability distribution (\ref{bolzatom}) is the Boltzmann-Gibbs distribution resulting from the free-system Hamiltonian $H_s(\bi{p})$ in (\ref{Hgas}). 
Note that we do not use the semi-classical phase-space volume element $\rmd^3q\,\rmd^3p/(2\pi\hbar)^3$, as our model is purely classical. The Helmholtz free energy of the particle is thus
\begin{equation}    \label{Fatom}
F=-k_BT\ln Z=-k_BT \ln\left[V\left(2\pi mk_BT\right)^{3/2}\right].
\end{equation}

Because of the coupling to the environment ($H_\mathrm{int}$ in (\ref{Hgas})), the free energy (\ref{Fatom}) is not the quantity $F^\star$ that appears in the Jarzynski equality (\ref{jar}). The Hamiltonian of mean force (\ref{Hmf}) has a contribution from the interaction term $H_\mathrm{int}$ in (\ref{Hgas}), and this gives a contribution from the interaction to $Z^\star$ and $F^\star$ (equations (\ref{Zstar}) and (\ref{Fstar})). We do not attempt to calculate $F^\star$ here, but return later to its relevance for the Jarzynski equality.

To explore the Jarzynski equality in this model, we consider a thought experiment. We drive the particle of interest out of equilibrium with the rest of the gas by continuously changing its mass over a time interval $0\leq t\leq\tau$ according to
\begin{equation}  \label{mtau}
m_t=m+\frac{t}{\tau}(m_\tau-m).
\end{equation}
This non-equilibrium process is covered by the Jarzynski equality since it corresponds to a continuous change of a parameter ($m$) in the system Hamiltonian $H_s$ that does not appear in the interaction $H_\mathrm{int}$ with the environment~\cite{jar04}. The work done on the particle in this process would be trivial if the particle were a free system undergoing Hamiltonian evolution determined by $H_s$---in that case the work done would be just the change in energy of the particle. Collisions with the other particles during the process described by (\ref{mtau}) make the calculation of the work more interesting.

\section{Work done along trajectories with and without a collision}
If the particle of interest moves freely between collisions (without its mass changing), its momentum $\bi{p}$ is constant and its energy is given by $H_s=p^2/(2m)$. The momentum between collisions is constant even when the mass of the particle is changing continuously in time according to (\ref{mtau}) because we have defined this process as a change of the mass in the system \emph{Hamiltonian} (as opposed to, say, the Lagrangian). With the changing mass parameter the system Hamiltonian is $H_s=p^2/(2m_t)$, and so $\dot{\bi{p}}=-\partial H_s/\partial\bi{q}=0$. The Jarzynski equality only applies to parameter changes in the Hamiltonian; thus a process in which the particle Lagrangian is $L_s=m_tv^2/2$, with $m_t$ given by (\ref{mtau}), would change the momentum of the driven particle and give a work relation that differs from the Jarzynski equality.

\begin{figure}[!htbp]
\begin{center} 
\includegraphics[width=10cm]{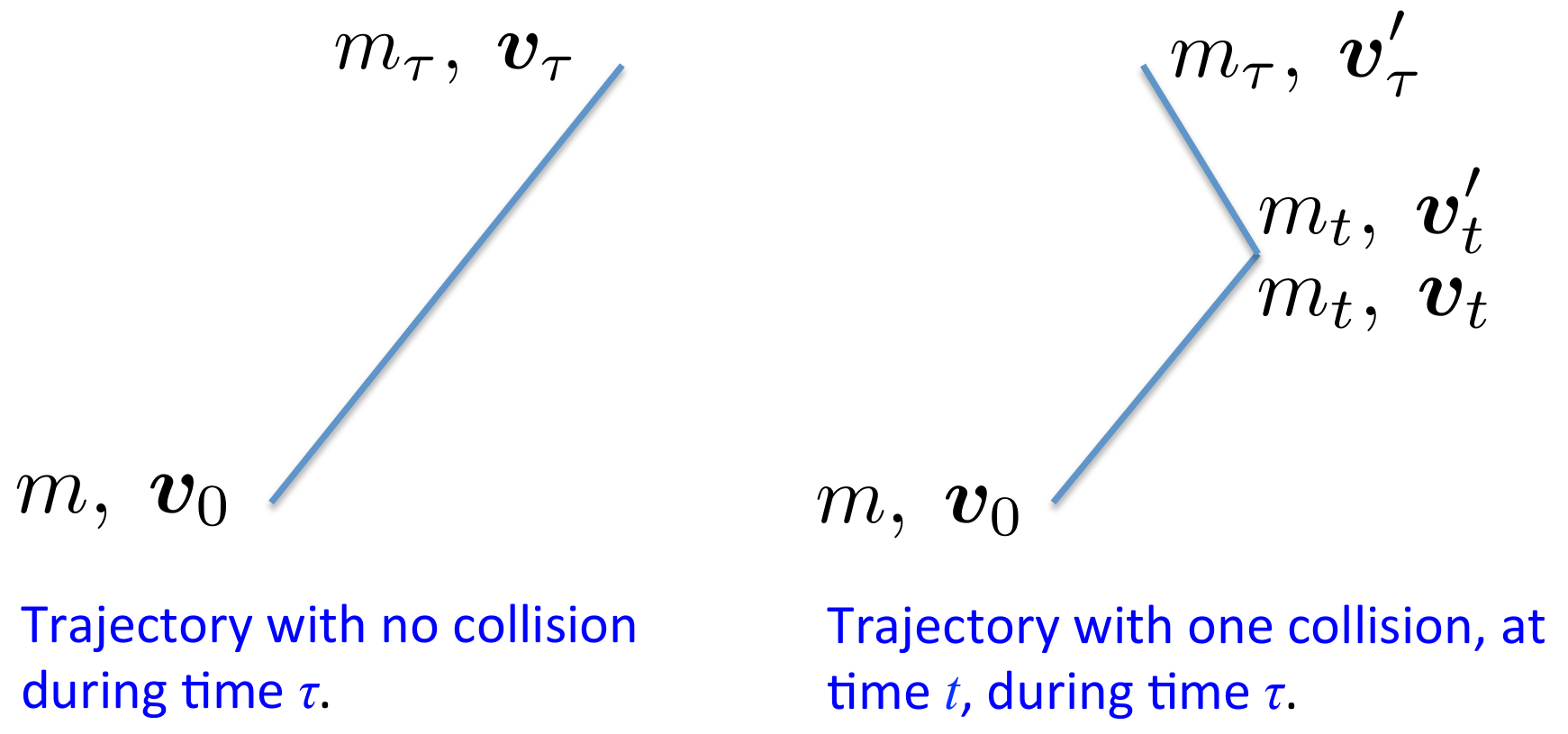}
\caption{Trajectories of the driven particle in the time interval $0\leq t\leq\tau$ with no collision (left) and with one collision at time $t$ (right). As the momentum of the particle is conserved between collisions, while its mass $m_t$ changes, its velocity $\bi{v}_t$ changes with time. A collision at time $t$ (right) changes the velocity discontinuously from $\bi{v}_t$ to a new value $\bi{v}'_t$.}
\label{fig:coll}
\end{center}
\end{figure}

A collision changes the momentum of the driven particle discontinuously to a new value that is again conserved until the next collision. During the time interval $\tau$ in which the mass of the particle is given by (\ref{mtau}), the Maxwell distribution of velocities implies a non-zero probability of any number of collisions occurring. Larger numbers of collisions are more unlikely than smaller numbers, however, and to make the problem tractable we will choose $\tau$ small enough so that it is highly probable there will be no collisions during this time interval, with a very small probability of one collision. The probabilities of two or more collisions during the time $\tau$ will be neglected as negligible. This approximation will not prevent us from checking the Jarzynski relation (\ref{jar}) as a strict equality because the inclusion of higher numbers of collisions leads to terms with higher powers of $\tau$ and the equality (\ref{jar}) must separately hold for all orders of $\tau$. We thus consider only two types of trajectories for the driven particle in the time interval $\tau$: trajectories with no collisions and trajectories with one collision (see Fig.~\ref{fig:coll}).

Let the velocity of the particle of interest at $t=0$ be $\bi{v}_0$. The momentum of the particle is conserved for $t>0$ as long as there is no collision, while its mass $m_t$ changes; hence its velocity $\bi{v}_t$ at time $t>0$ satisfies
\begin{equation}  \label{momcon}
m\bi{v}_0=m_t\bi{v}_t
\end{equation}
if there is no collision. For the first type of trajectory, which has no collision for the entire interval $\tau$, the work done on the particle throughout the change of its mass is the resulting change of energy (with constant momentum $m\bi{v}_0$):
\begin{equation} \label{w1}
w_\mathrm{free}=\frac{1}{2}m^2v_0^2\left(\frac{1}{m_\tau}-\frac{1}{m}\right).
\end{equation}
The second  type of trajectory has a collision at time $t$ during the interval $\tau$, in which the velocity of the particle is changed instantly from $\bi{v}_t$ to a new value $\bi{v}'_t$. The work done on the particle in this trajectory~\cite{Crooks98, Sagawa10} is the sum of the change in its energy in the interval $0$ to $t$ before the collision (with constant momentum $m\bi{v}_0$) and the change in its energy in the interval $t$ to $\tau$ after the collision (with constant momentum $m_t\bi{v}'_t$):
\begin{equation} \label{w2}
w_\mathrm{col}=\frac{1}{2}m^2v_0^2\left(\frac{1}{m_t}-\frac{1}{m}\right)+\frac{1}{2}m_t^2{v_t'}^2\left(\frac{1}{m_\tau}-\frac{1}{m_t}\right).
\end{equation}

To calculate averages involving the work done on the driven particle, we compute separately the contributions of the two types of trajectories. For trajectories of the first type we must weight the average with the probability of no collision occurring during the interval $\tau$, and also average over the initial velocity $\bi{v}_0$ of the particle. For trajectories of the second type we must weight averages with the probability of a collision at time $t$ giving the particle a velocity $\bi{v}'_t$, average over the possible post-collision velocities $\bi{v}'_t$, sum (integrate) over all collision times $t$ in the interval $0\leq t\leq\tau$, and average over the initial velocity $\bi{v}_0$.

\section{Collision statistics}
The statistics of a collision is the only challenging aspect of the model. While the collision \emph{rate} in a perfect gas is not difficult to obtain~\cite{LLsp1}, we require something much more detailed for the calculation of the average work and average exponentiated work for the mass variation process. Specifically, we need the statistics of the final velocity of a colliding particle that has been driven out of equilibrium with the rest of the gas (the work (\ref{w2}) contains the post-collision speed $v'_t$ of the driven particle).

\begin{figure}[!htbp]
\begin{center} 
\includegraphics[width=6cm]{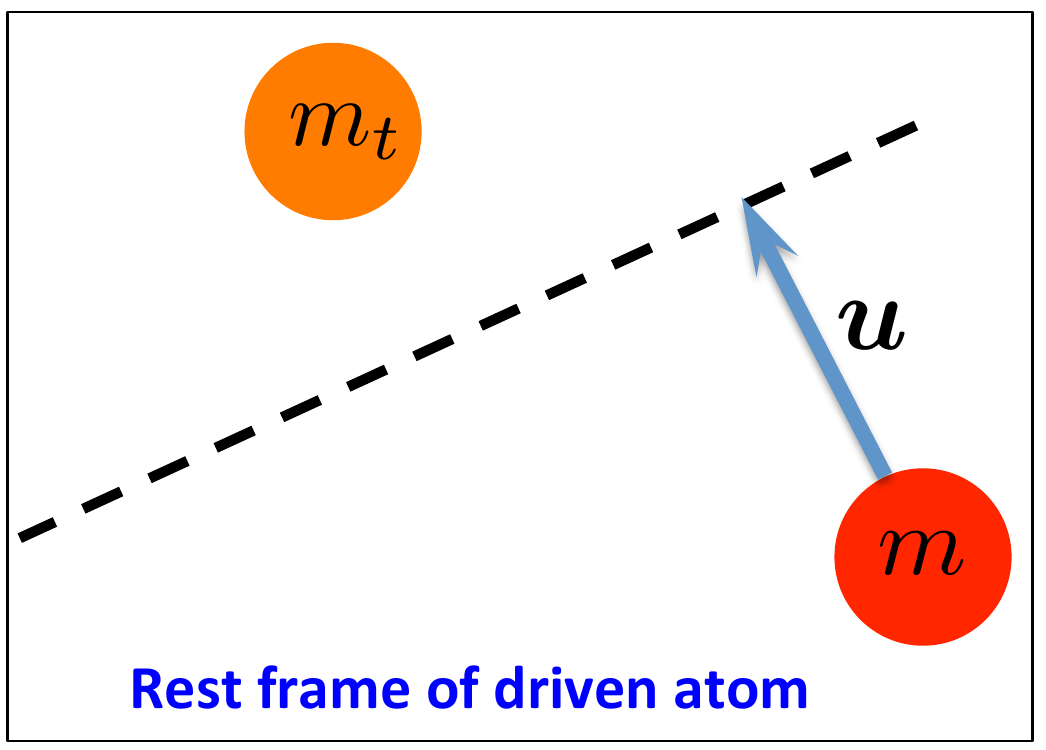}
\caption{In the rest frame of the driven particle (with mass $m_t$), another particle (with mass $m$) moves at the relative velocity $\bi{u}$, covering a distance $u=|\bi{u}|$ per unit time.}
\label{fig:flux}
\end{center}
\end{figure}

The probability of the driven particle colliding with another particle in unit time is proportional to the flux of the other particles in the reference frame of the driven particle. 
As the other particles are in thermal equilibrium at temperature $T$, their velocities are Maxwell distributed, with each particle having a probability $\rho(v) \rmd^3v$ of moving at a velocity between $\bi{v}$ and $\bi{v}+\rmd\bi{v}$, where~\cite{LLsp1}
\begin{equation}  \label{vdis}
\rho(v) \rmd^3v=\left(\frac{m}{2\pi k_BT}\right)^{3/2}\exp\left(-\frac{mv^2}{2k_BT}\right)\rmd^3v.
\end{equation}
The thermal distributions of the momenta of the particles are unaffected by the interaction terms in the Hamiltonian (\ref{Hgas}) because the latter do not contain the momenta. The joint velocity distribution of the particles is therefore just the product of Maxwell distributions (\ref{vdis}). (The joint position distributions of the particles will, however, depend on the interaction; in the simple case of potential barriers the joint distribution will rule out spatial overlaps of the particles.) In the rest frame of the driven particle (Fig.~\ref{fig:flux}) a particle with velocity $\bi{v}$ in the laboratory frame moves with the relative velocity
\begin{equation}  \label{u}
\bi{u}=\bi{v}-\bi{v}_t,
\end{equation}
where $\bi{v}_t$ is the velocity of the driven particle. The other particle moves a distance $u=|\bi{u}|$ per unit time in the frame of the driven particle. If there are $N$ other particles ($N+1$ particles in total) then the flux of other particles with relative velocity between $\bi{u}$ and $\bi{u}+\rmd \bi{u}$ incident on the driven particle is $(N/V) u\, \rho(v) \rmd^3v$, where we must substitute for $v$ in terms of $u$ using (\ref{u}). Defining $\theta$ as the angle between the relative velocity $\bi{u}$ and the velocity $\bi{v}_t$ of the driven particle, this flux is, from (\ref{vdis}) and (\ref{u}),
\begin{equation}  \label{flux}
\fl
\frac{N}{V}\left(\frac{m}{2\pi k_BT}\right)^{3/2}\exp\left[-\frac{m}{2k_BT}\left(u^2+2uv_t\cos\theta+v_t^2\right)\right]u^32\pi\sin\theta\,\rmd\theta\,\rmd u.
\end{equation}

The probability per unit time of the driven particle colliding with a particle having relative velocity between $\bi{u}$ and $\bi{u}+\rmd \bi{u}$ is proportional to the flux (\ref{flux}). Denoting this probability per unit time by $P(u,\theta,v_t)\rmd\theta\,\rmd u$, the probability for such a collision to occur between times $t$ and $t+\rmd t$ is $P(u,\theta,v_t)\rmd\theta\,\rmd u\,\rmd t$, and the probability for such a collision \emph{not} to occur between times $t$ and $t+\rmd t$ is $1-P(u,\theta,v_t)\rmd\theta\,\rmd u\,\rmd t$. The probability of one collision occurring between times $t$ and $t+\rmd t$ during the time interval $\tau$ is thus
\begin{eqnarray}  
\fl
\left(1-\int_0^t \rmd t'\,P(u',\theta',v_{t'})\rmd\theta'\,\rmd u'\right)P(u,\theta,v_t)\rmd\theta\,\rmd u\,\rmd t    \nonumber \\
   \qquad\qquad\qquad \qquad \times \left(1-\int_t^\tau \rmd t''\,P(u'',\theta'',v_{t''})\rmd\theta''\,\rmd u''\right),   \label{pcol}
\end{eqnarray}
and we recall that the collision is with a particle having relative velocity between $\bi{u}$ and $\bi{u}+\rmd \bi{u}$. Trajectories with one collision during the interval $\tau$ have a probability that contains (\ref{pcol}), averaged over the time $t$ of the collision. As described in the last section, we take the time interval $\tau$ to be so short that the probability of a collision during this time is very small. It follows that we may use $P(u,\theta,v_t)\rmd\theta\,\rmd u\,\rmd t$, instead of (\ref{pcol}), in treating trajectories with one collision, which implies our results for these trajectories will be first order in $\tau$. Analysis of trajectories with more than one collision will necessarily lead to higher-order terms in $\tau$ and the Jarzynski equality must hold separately for each order of $\tau$.

\begin{figure}[!htbp]
\begin{center} 
\includegraphics[width=9cm]{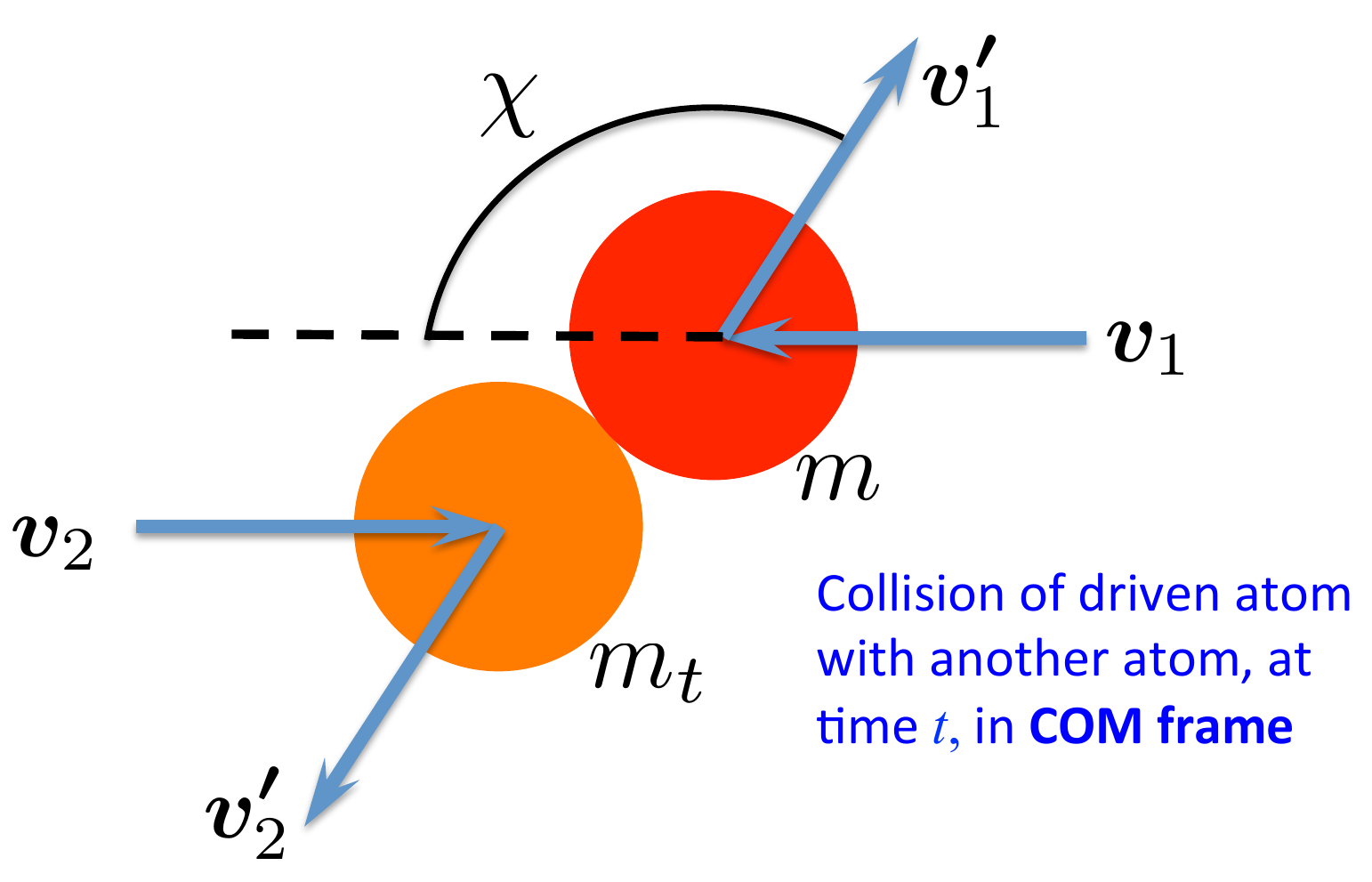}
\caption{Collision of the driven particle (mass $m_t$) with another particle (mass $m$) at time $t$ in the centre of mass (COM) frame of the two particles.}
\label{fig:com}
\end{center}
\end{figure}

The probability per unit time of the driven particle colliding is the flux of incident particles times the scattering cross section. We will write the differential scattering cross section in terms of the scattering angle $\chi$ in the centre of mass (COM) frame (Fig.~\ref{fig:com}). In the COM frame~\cite{LLmec}, the velocities of the two particles before the collision are in opposite directions, as are the velocities after the collision, and the collision only changes the direction of each velocity, not their magnitudes. We denote, in the COM frame, the pre-collision velocities of the driven particle and the particle with which it collides by $\bi{v}_2$ and $\bi{v}_1$, respectively (Fig.~\ref{fig:com}). The corresponding post-collision velocities are denoted $\bi{v}'_2$ and $\bi{v}'_1$. We then have~\cite{LLmec}
\begin{eqnarray}
\bi{v}_1=\frac{m_t}{m+m_t}\,\bi{u}, \qquad \bi{v}_2=-\frac{m}{m+m_t}\,\bi{u} ,  \label{v12} \\
|\bi{v}'_1|=|\bi{v}_1|, \qquad |\bi{v}'_2|=|\bi{v}_2| , \qquad \bi{v}_1\boldsymbol{\cdot}\bi{v}'_1=v_1^2\cos\chi,  \label{v12p}
\end{eqnarray}
where $\bi{u}$ is the pre-collision relative velocity (\ref{u}) in the laboratory frame (which is also the pre-collision relative velocity in the COM frame: $\bi{u}=\bi{v}_1-\bi{v}_2$) and $\chi$ is the scattering angle in the COM frame. We have already denoted the post-collision velocity of the driven particle in the laboratory frame by $\bi{v}'_t$; the post-collision velocity of the other particle in the laboratory frame is denoted $\bi{v}'$. In terms of the pre-collision laboratory frame velocities ($\bi{v}$ and $\bi{v}_t$) and the post-collision COM-frame velocity $\bi{v}'_1$ of the non-driven particle, $\bi{v}'$ and $\bi{v}'_t$ can be written~\cite{LLmec}
\begin{equation}  \label{vspost}
\bi{v}'=\bi{v}'_1+\frac{m\bi{v}+m_t\bi{v}_t}{m+m_t}, \qquad  \bi{v}'_t=-\frac{m}{m_t}\bi{v}'_1+\frac{m\bi{v}+m_t\bi{v}_t}{m+m_t}.
\end{equation}
Recalling that the particles are spheres of radius $R$, the impact parameter is $2R\cos(\chi/2)$ and the differential cross section $\rmd\sigma$ is~\cite{LLmec}
\begin{equation}  \label{dsig}
\rmd\sigma=\frac{1}{4}(2R)^2\sin\chi\,\rmd\chi\,\rmd\phi,
\end{equation}
where $\phi$ is the azimuthal angle in the COM frame corresponding to rotating the post-collision velocities for fixed scattering angle $\chi$ around the pre-collision velocities. 

The probability of a particle with relative velocity between $\bi{u}$ and $\bi{u}+\rmd \bi{u}$ colliding with the driven particle at a time between $t$ and $t+\rmd t$ (where $t$ is less than the very short time interval $\tau$) and being scattered into the solid angle element $\rmd\chi\,\rmd\phi$ centred on $(\chi,\phi)$ is equal to the flux (\ref{flux}) times $\rmd t$ times the differential cross section (\ref{dsig}):
\begin{equation}  \label{dpcol}
\fl
\frac{2\pi R^2 N}{V}\left(\frac{m}{2\pi k_BT}\right)^{3/2}\exp\left[-\frac{m}{2k_BT}\left(u^2+2uv_t\cos\theta+v_t^2\right)\right]u^3 \sin\theta\,\sin\chi\,\rmd u\,\rmd\theta\,\rmd\chi\,\rmd\phi\,\rmd t.
\end{equation}
This is the (differential) probability for a single trajectory of the driven particle, with initial velocity $\bi{v}_0$, that includes one collision. The (very small) total probability of the driven particle with initial velocity $\bi{v}_0$ following a trajectory with one collision during the (very short) time interval $\tau$ is obtained by integrating (\ref{dpcol}) over all relative speeds $u$, all angles $\theta$ between $\bi{u}$ and $\bi{v}_t$, all scattering angles $(\chi,\phi)$, and all possible collision times $t$ in the interval $0\leq t \leq\tau$. For averages involving the work done on the driven particle in trajectories with a collision, we require the differential probability (\ref{dpcol}) of each trajectory with a collision since the work depends on the details of the trajectory (relative velocity, scattering angle and time of collision).

\begin{figure}[!htbp]
\begin{center} 
\includegraphics[width=10cm]{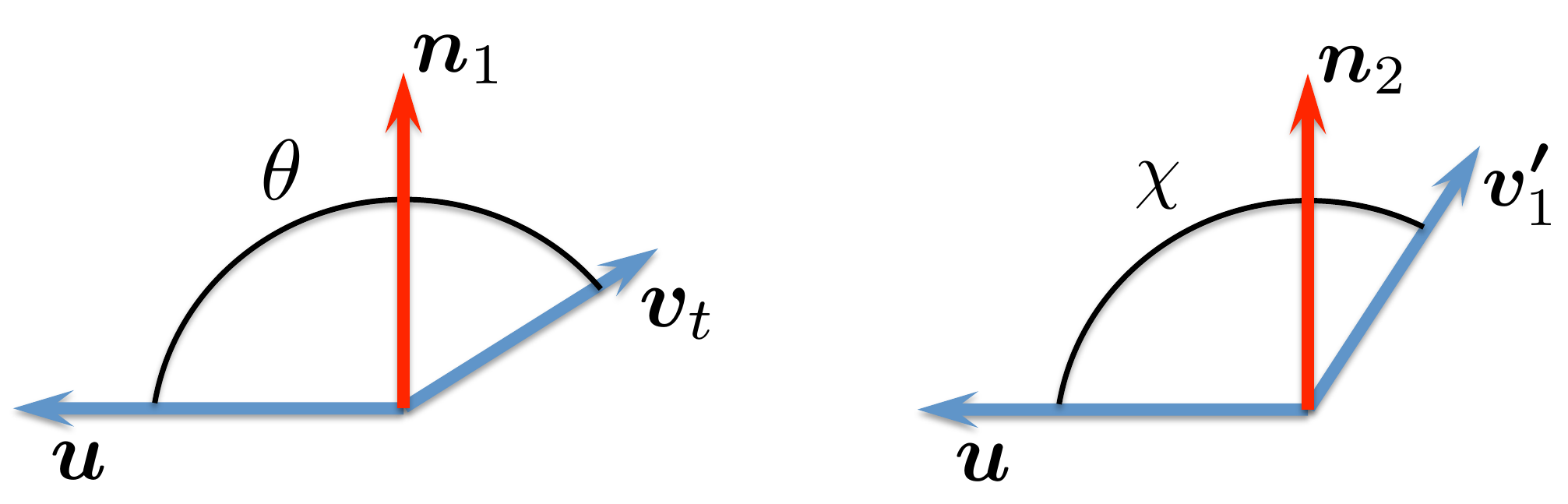}
\caption{Left: The plane of the relative velocity $\bi{u}$ of the colliding particles and the pre-collision velocity $\bi{v}_t$ of the driven particle. Right: The plane of the relative velocity $\bi{u}$ and the post-collision velocity $\bi{v}'_1$ in the COM frame of the non-driven particle.}
\label{fig:angles}
\end{center}
\end{figure}

The work done (\ref{w2}) on the driven particle in a trajectory with a collision features its post-collision speed $v'_t$; in order to calculate work averages it is necessary to express this speed in terms of the variables $u$, $\theta$, $\chi$ and $\phi$ appearing in the differential probability (\ref{dpcol}) for a trajectory with a collision. From the second of (\ref{vspost}) we obtain the following expression for ${v'_t}^2$:
\begin{equation}  \label{vtsq0}
\fl
{v'_t}^2=\frac{2m^2}{(m+m_t)^2}u^2+\frac{2m}{m+m_t}uv_t\cos\theta+v_t^2-\frac{2m^2}{m_t(m+m_t)}\bi{v}'_1\boldsymbol{\cdot}\bi{u}-\frac{2m}{m_t}\bi{v}'_1\boldsymbol{\cdot}\bi{v}_t,
\end{equation}
where we have used the definition of $\theta$, introduced before equation (\ref{flux}), as the angle between $\bi{u}$ and $\bi{v}_t$. The relative velocity $\bi{u}$ is in the direction of $\bi{v}_1$ (since $\bi{u}=\bi{v}_1-\bi{v}_2$), hence (see Fig.~\ref{fig:com}) the angle between $\bi{u}$ and $\bi{v}'_1$ is the scattering angle $\chi$ and we have
\begin{equation}  \label{v1pu}
\bi{v}'_1\boldsymbol{\cdot}\bi{u}=v'_1u\cos\chi=v_1u\cos\chi=\frac{m_t}{m+m_t}u^2\cos\chi,
\end{equation}
where we have used (\ref{v12p}) and (\ref{v12}). To simplify the dot product $\bi{v}'_1\boldsymbol{\cdot}\bi{v}_t$ in (\ref{vtsq0}), we refer to Fig.~\ref{fig:angles}. Recall that $\theta$ is the angle between $\bi{u}$ and $\bi{v}_t$ (left of Fig.~\ref{fig:angles}) and introduce a unit vector $\bi{n}_1$ orthogonal to $\bi{u}$, lying in the plane of $\bi{u}$ and $\bi{v}_t$. We can then expand $\bi{v}_t$ as
\begin{equation} \label{vtexp}
\bi{v}_t=\frac{\bi{u}}{u}v_t\cos\theta+\bi{n}_1v_t\sin\theta.
\end{equation}
As noted above equation (\ref{v1pu}), $\chi$ is the angle between $\bi{u}$ and $\bi{v}'_1$ (right of Fig.~\ref{fig:angles}). Introducing a unit vector $\bi{n}_2$ orthogonal to $\bi{u}$, lying in the plane of $\bi{u}$ and $\bi{v}'_1$, we can expand $\bi{v}'_1$ as
\begin{eqnarray}
\bi{v}'_1&=\frac{\bi{u}}{u}v'_1\cos\chi+\bi{n}_2v'_1\sin\chi=\frac{\bi{u}}{u}v_1\cos\chi+\bi{n}_2v_1\sin\chi   \nonumber \\
&=\frac{m_t}{m+m_t}u\left(\frac{\bi{u}}{u}\cos\chi+\bi{n}_2\sin\chi   \right),  \label{v1pexp}
\end{eqnarray}
where we used (\ref{v12p}) and (\ref{v12}). The unit vectors $\bi{n}_1$ and $\bi{n}_2$ lie in a plane orthogonal to the relative velocity $\bi{u}(=\bi{v}_1-\bi{v}_2)$, and hence they lie in a plane orthogonal to the pre-collision velocities $\bi{v}_1$ and $\bi{v}_2$ in the COM frame (see Fig.~\ref{fig:com}). We can therefore define the angle $\phi$ in the differential cross section (\ref{dsig}), which measures a rotation of the post-collision velocities about the line of the pre-collision velocities in the COM frame, as the angle between the unit vectors $\bi{n}_1$ and $\bi{n}_2$. The expansions (\ref{vtexp}) and (\ref{v1pexp}) then give the dot product
\begin{equation}  \label{v1pvt}
\bi{v}'_1\boldsymbol{\cdot}\bi{v}_t=\frac{m_t}{m+m_t}uv_t\left(\cos\chi\cos\theta+\sin\chi\sin\theta\cos\phi\right).
\end{equation}
Inserting (\ref{v1pu}) and (\ref{v1pvt}) in (\ref{vtsq0}), we find the following expression for the square of the post-collision speed of the driven particle in the laboratory frame:
\begin{equation}  \label{vtsq}
\fl
{v'_t}^2=v_t^2+\frac{2m^2}{(m+m_t)^2}u^2\left(1-\cos\chi\right)+\frac{2m}{m+m_t}uv_t\left[ \cos\theta\left(1-\cos\chi\right)-\sin\chi\sin\theta\cos\phi\right].
\end{equation}

\section{Work averages}
We now have the ingredients to compute work averages for the driven particle. In our approximation of a very short driving time $\tau$, these averages have contributions from trajectories with no collision and from trajectories with one collision. In order to analytically evaluate the integrals involved we must make an expansion in terms of $m_\tau-m$, the total change in mass of the driven particle during the driving time $\tau$. While it is in principle possible to expand the results to arbitrary order in $m_\tau-m$, we will only do so up to order $(m_\tau-m)^2$. As with expansions in the driving time $\tau$, any strict equalities involving work averages must hold separately for all orders of $m_\tau-m$. The quantities $m_t$ and $v_t$ that appear in the integrands must be written in terms of $m$ and $m_\tau$ in order to expand in powers of $m_\tau-m$; we express $m_t$ in terms of $m$ and $m_\tau$ using (\ref{mtau}) and $v_t$ in expressed in terms of $v_0$, $m$ and $m_\tau$ using (\ref{momcon}) and (\ref{mtau}). 

\subsection{Work averages for trajectories with no collision}
The work done on the driven particle, with initial velocity $\bi{v}_0$, in a trajectory with no collision is given by (\ref{w1}). Averages over such trajectories of functions of the work (\ref{w1}) must be weighted with the probability of no collision occurring during the driving time $\tau$. As noted after (\ref{dpcol}), the probability of a collision occurring during the driving time $\tau$, for initial  velocity $\bi{v}_0$ of the driven particle, is given by integrating (\ref{dpcol}) over all trajectories with a collision; we denote this probability by  $P_\mathrm{col}(\tau, v_0)$. Expanding (\ref{dpcol}) to order $(m_\tau-m)^2$ and integrating over $\phi$, $\chi$, $\theta$, $u$ (from $0$ to $\infty$), and $t$ (from $0$ to $\tau$) gives
\begin{eqnarray}
\fl
P_\mathrm{col}(\tau, v_0)=\frac{4R^2\tau}{mv_0}\left[\sqrt{2\pi mk_BT}\,v_0\exp\left(-\frac{mv_0^2}{2k_BT}\right)+\pi\left(k_BT+mv_0^2\right)\mathrm{erf}\left(v_0\sqrt{\frac{m}{2k_BT}}\right)  \right] \nonumber \\
\fl
\quad - \frac{2R^2\tau}{m^2v_0}\left[\sqrt{2\pi mk_BT}\,v_0\exp\left(-\frac{mv_0^2}{2k_BT}\right)-\pi\left(k_BT-mv_0^2\right)\mathrm{erf}\left(v_0\sqrt{\frac{m}{2k_BT}}\right)  \right] (m_\tau-m) \nonumber \\
\fl
\quad +\frac{4\pi R^2v_0\tau}{3m^2}\,\mathrm{erf}\left(v_0\sqrt{\frac{m}{2k_BT}}\right)(m_\tau-m)^2+O(m_\tau-m)^3,  \label{probcol}
\end{eqnarray}
where $\mathrm{erf}(x)=(2/\sqrt{\pi})\int_0^x\rmd t\,e^{-t^2}$ is the error function. This is the probability of the driven particle, with initial velocity $\bi{v}_0$, experiencing a collision during the driving time $\tau$. The probability of no collision, for initial velocity $\bi{v}_0$, during the time $\tau$ is one minus (\ref{probcol}).

Consider the exponentiated work average $\left\langle \exp\left[- w/(k_BT)\right] \right\rangle$ that appears in the Jarzynski equality (\ref{jar}). The contribution to $\left\langle \exp\left[- w/(k_BT)\right] \right\rangle$ from trajectories without a collision is obtained by taking $\exp\left[- w_\mathrm{free}/(k_BT)\right]$, with $w_\mathrm{free}$ given by (\ref{w1}),  multiplying by $(1-P_\mathrm{col}(\tau, v_0))$ and averaging over the initial velocity $\bi{v}_0$ of the driven particle. We denote this contribution to $\left\langle \exp\left[- w/(k_BT)\right] \right\rangle$ by $\left\langle \exp\left[- w_\mathrm{free}/(k_BT)\right] \right\rangle_\mathrm{free}$. At $t=0$ the driven particle is still in thermal equilibrium with the rest of the gas so $\bi{v}_0$ is Maxwell distributed with probability distribution (\ref{vdis}). Performing the average over $\bi{v}_0$, in which we expand the integrand to second order in $m_\tau-m$, we obtain
\begin{eqnarray}
\fl
\left\langle \exp\left(- \frac{w_\mathrm{free}}{k_BT}\right) \right\rangle_{\!\!\!\mathrm{free}}=&1-16R^2\tau\frac{N}{V}\sqrt{\frac{\pi k_BT}{m}} +\left(\frac{3}{2}-24R^2\tau\frac{N}{V}\sqrt{\frac{\pi k_BT}{m}}\right)\frac{ m_\tau-m}{m}   \nonumber  \\
\fl
&+\left(\frac{9}{24}-\frac{35}{6}R^2\tau\frac{N}{V}\sqrt{\frac{\pi k_BT}{m}}\right)\frac{(m_\tau-m)^2}{m^2} +O(m_\tau-m)^3.    \label{expwfree} 
\end{eqnarray}

For consideration of the inequality (\ref{wDF}), we require the average work $\left\langle w \right\rangle$. The contribution to $\left\langle w \right\rangle$ from trajectories without a collision, which we denote by $\left\langle w_\mathrm{free} \right\rangle_\mathrm{free}$, is computed as described in the last paragraph, with $w_\mathrm{free}$ replacing $\exp\left[- w_\mathrm{free}/(k_BT)\right]$. The result is
\begin{eqnarray}
\fl
\left\langle w_\mathrm{free}  \right\rangle_{\mathrm{free}}=&-\left[\frac{3k_BT}{2m}-28\sqrt{\pi}R^2\tau\frac{N}{V}\left(\frac{k_BT}{m}\right)^{3/2}\right] (m_\tau-m)   \nonumber  \\
\fl
&+\left[\frac{3k_BT}{2m}-37\sqrt{\pi}R^2\tau\frac{N}{V}\left(\frac{k_BT}{m}\right)^{3/2}\right]\frac{(m_\tau-m)^2}{m} +O(m_\tau-m)^3.    \label{wfree} 
\end{eqnarray}

\subsection{Work averages for trajectories with a collision}
The work done on the driven particle, with initial velocity $\bi{v}_0$, in a trajectory with a collision is given by (\ref{w2}). Averages over such trajectories of functions of the work (\ref{w2}) must be weighted with the probability of each trajectory, and these probabilities are given by (\ref{dpcol}) with the values of $u$, $\theta$, $\chi$, $\phi$ and $t$ (collision time) that characterise the trajectory. The work (\ref{w2}) also depends on the trajectory through the same set of parameters, as can be seen by substituting (\ref{vtsq}) into (\ref{w2}).

The contribution of trajectories with a collision to the average $\left\langle \exp\left[- w/(k_BT)\right] \right\rangle$ is denoted by $\left\langle \exp\left[- w_\mathrm{col}/(k_BT)\right] \right\rangle_\mathrm{col}$. We compute this by weighting $ \exp\left[- w_\mathrm{col}/(k_BT)\right]$ with the probability (\ref{dpcol}) of the trajectory corresponding to $w_\mathrm{col}$ and integrating over all trajectories with a collision. Before performing the integrations we expand the integrand to second order in $m_\tau-m$. The integrations are over $u$, $\theta$, $\chi$, $\phi$ and $t$ and the result is
\begin{eqnarray}
\fl
\left\langle \exp\left(- \frac{w_\mathrm{col}}{k_BT}\right) \right\rangle_{\!\!\!\mathrm{col}}=&R^2\tau\frac{N}{V}\sqrt{\frac{\pi k_BT}{m}}\left[16+\frac{24}{m}(m_\tau-m)+\frac{35}{6m^2}(m_\tau-m)^2\right]   \nonumber  \\
\fl
&+O(m_\tau-m)^3.    \label{expwcol} 
\end{eqnarray}

The contribution of trajectories with a collision to the average work $\left\langle  w \right\rangle$ is denoted by $\left\langle w_\mathrm{col} \right\rangle_\mathrm{col}$. This is calculated in the same manner as described in the last paragraph and the result is
\begin{eqnarray}
\fl
\left\langle w_\mathrm{col}  \right\rangle_{\mathrm{col}}=\sqrt{\pi}R^2\tau\frac{N}{V}\left(\frac{k_BT}{m}\right)^{3/2}\left[-28(m_\tau-m)+\frac{103}{3m}(m_\tau-m)^2\right]   +O(m_\tau-m)^3.    \label{wcol} 
\end{eqnarray}

\section{Jarzynski equality and the second law}
By adding the two contributions (\ref{expwfree}) and (\ref{expwcol}), we find the average $\left\langle \exp\left[- w/(k_BT)\right] \right\rangle$ of the exponentiated work done on the driven particle, to order $(m_\tau-m)^2$:
 \begin{equation}
\fl
\left\langle \exp\left(- \frac{w}{k_BT}\right) \right\rangle=1+\frac{3}{2m}(m_\tau-m)+\frac{3}{8m^2}(m_\tau-m)^2  +O(m_\tau-m)^3.   \label{expw} 
\end{equation}
Note that the effect of the collisions, which shows up in terms containing (among other quantities) the particle radius $R$, has entirely cancelled out in (\ref{expw}). The \emph{equilibrium} Helmholtz free-energy difference $\Delta F$ of the driven particle between the equilibrium states corresponding to the final system parameter (mass $m_\tau$) and the initial system parameter (mass $m$) is shown by (\ref{Fatom}) to be $\Delta F=-(3/2)k_BT\ln(m_\tau/m)$, which gives
\begin{equation}
\fl
\exp\left(-\frac{\Delta F}{k_BT}\right)=\left(\frac{m_\tau}{m}\right)^{3/2}=1+\frac{3}{2m}(m_\tau-m)+\frac{3}{8m^2}(m_\tau-m)^2  +O(m_\tau-m)^3.  \label{expDF}
\end{equation}
The results (\ref{expw}) and (\ref{expDF}) indicate that the Jarzynski equality holds for this process in the form (\ref{jar2}), despite the fact that the system is coupled to the environment. In fact there is no reason why (\ref{jar2}) should hold only up to the accuracy of our calculation and cease to hold for higher orders in $m_\tau-m$ and $\tau$. In the next section we show that the process we have considered is such that $\Delta F^\star=\Delta F$, so we have found a case where the Helmholtz free-energy $F$ of the system gives the exact work relation (\ref{jar2}) even though $F$ neglects the coupling to the environment.

The formulation (\ref{wDF}) of the second law also follows from the Jarzynski equality in the form (\ref{jar2})~\cite{jar97}. For systems coupled to the environment, however, the Jarzynski equality (in general) takes the form (\ref{jar}), leading to the inequality $\langle w \rangle \geq \Delta F^\star$. In our model the Jarzynski equality takes the form (\ref{jar2}), despite the coupling to the environment (see next section for a complete proof), and thus the inequality $\langle w \rangle \geq \Delta F$ must strictly hold. Adding (\ref{wfree}) and (\ref{wcol}), we obtain the average work
 \begin{equation}
\fl
\left\langle w \right\rangle=-\frac{3k_BT}{2m} (m_\tau-m)  
+\left[\frac{3k_BT}{2m}-\frac{8}{3}\sqrt{\pi}R^2\tau\frac{N}{V}\left(\frac{k_BT}{m}\right)^{3/2}\right]\frac{(m_\tau-m)^2}{m} +O(m_\tau-m)^3.    \label{w} 
\end{equation}
In contrast to (\ref{expw}), the average work (\ref{w}) does depend on the collisions (coupling to the environment). The equilibrium free-energy difference $\Delta F=-(3/2)k_BT\ln(m_\tau/m)$ gives
\begin{equation}
\Delta F=k_BT\left[-\frac{3}{2m} (m_\tau-m) +\frac{3}{4m^2} (m_\tau-m)^2 \right]   +O(m_\tau-m)^3.  \label{DF}
\end{equation}
Comparing (\ref{w}) and (\ref{DF}), we find the inequality  $\langle w \rangle \geq \Delta F$ requires
\begin{equation}   \label{tauineq}
\frac{3}{16}\geq R^2\tau\frac{N}{V}\sqrt{\frac{\pi k_BT}{m}}+O(m_\tau-m)^3.
\end{equation}
Our calculation is only valid for very small $\tau$ such that we can neglect terms of order $\tau^2$ and higher; the bound on $\tau$ given by (\ref{tauineq}) can thus be met by our approximations. The results of the next section show that $\langle w \rangle \geq \Delta F$ will also hold for our model if the exact average work $\langle w \rangle $ is calculated.

\section{General condition for cancellation of environment coupling in the Jarzynski equality}
For a system coupled to the environment, the Jarzynski equality is (\ref{jar2}), where the work is performed by a change in a parameter of the system Hamiltonian $H_s(x)$. Denote the work parameter (which was the particle mass $m$ in our model) by $\lambda$. We can then include the work parameter in the functional dependence of the system Hamiltonian, so that it is denoted $H_s(x,\lambda)$. Now suppose $H_s(x,\lambda)$ can be written
\begin{equation}   \label{Hsdecom}
H_s(x,\lambda)=H^{(1)}_s(x_1,\lambda)+H^{(2)}_s(x_2),
\end{equation}
where $x_1$ and $x_2$ are disjoint subsets of the set of canonical variables $x=\{q_i,p_i\}$ of the system such that $x_1\cup x_2=x$. The decomposition (\ref{Hsdecom}) applies when there are canonical variables ($x_2$) of the system that do not appear in any of the terms of $H_s(x,\lambda)$ containing the work parameter $\lambda$. Suppose further that the interaction term $H_\mathrm{int}(x,X)$ in the full system-environment Hamiltonian (\ref{H}) only contains the subset $x_2$ of the system canonical variables, i.e.\
\begin{equation}  \label{Hintspec}
H_\mathrm{int}(x,X)=H_\mathrm{int}(x_2,X).
\end{equation}
Now consider the partition function (\ref{Zstar}), which we here denote by $Z^\star(\lambda)$ to record its dependence on $\lambda$. Using the definition (\ref{Hmf}) of the Hamiltonian of mean force, $Z^\star(\lambda)$ for the special case (\ref{Hsdecom}) and (\ref{Hintspec}) takes the form
\begin{eqnarray}  
\fl
Z^\star(\lambda)=&\int dx_1\,\exp\left[-H^{(1)}_s(x_1,\lambda)/(k_BT)\right] \nonumber \\
\fl 
&\times \frac{\int dx_2\,\int dX\,\exp\left\{-\left[H^{(2)}_s(x_2)+H_e(X)+H_\mathrm{int}(x_2,X)\right]/(k_BT)\right\}}{\int dX\,\exp\left[-H_e(X)/(k_BT)\right]}.  \label{Zstarspec}
\end{eqnarray}
It is clear from (\ref{Zstarspec}) that the ratio of the partition function $Z^\star(\lambda)$ for two different values $\lambda_A$ and $\lambda_B$ of the work parameter $\lambda$ is independent of the coupling to the environment, i.e.\ independent of $H_\mathrm{int}$:
\begin{equation}
\frac{Z^\star(\lambda_B)}{Z^\star(\lambda_A)}=\frac{\int dx_1\,\exp\left[-H^{(1)}_s(x_1,\lambda_B)/(k_BT)\right]}{\int dx_1\,\exp\left[-H^{(1)}_s(x_1,\lambda_A)/(k_BT)\right]}.   \label{Zstarratio}
\end{equation}
The difference $\Delta F^\star=F^\star(\lambda_B)-F^\star(\lambda_A)$ between the two values of the free energy (\ref{Fstar}) corresponding to $\lambda_A$ and $\lambda_B$ is determined by the ratio (\ref{Zstarratio}). Hence in this case the free-energy difference $\Delta F^\star$ is equal to the Helmholtz free-energy difference $\Delta F$ that ignores the coupling to the environment. This result establishes the following theorem regarding the Jarzynski equality: \emph{If the terms in the system Hamiltonian that contain the work parameter are independent of the system canonical variables that appear in the interaction with the environment, then the Jarzynski equality holds exactly with the uncoupled system free-energy difference $\Delta F$.} Note that this theorem is independent of the strength of the coupling to the environment.

We see from (\ref{Hgas}) that the above theorem applies to our model, and this is consistent with our results for the Jarzynski equality in the previous section.

\section{Conclusions}
Jarzynski's recent results~\cite{jar97,jar04} strengthen standard thermodynamics by providing an equality for a wide range of non-equilibrium processes. Our goal here has been to provide a model that is solvable and that includes coupling to an environment, in the belief that such models will help assess the significance of Jarzynski's results~\cite{jar97,jar04}. We have explored how Jarzynski's work equality  can be applied to one of the most basic thermodynamic systems---the perfect gas. The work done on a gas particle undergoing a change of mass was calculated taking into account the collisions of the particle with the gas. While the effect of collisions may be incorporated into the free energy of the particle, we have shown that its free energy difference is identical to an isolated particle and the standard Jarzynski equation holds. This is despite the coupling to the bath of other particles. Our model provides an example of an interesting general situation where (arbitrarily strong) coupling to the environment has no effect on Jarzynski's work relation. Given that the particular calculations presented here reveal a general theorem, the further study of simple models with coupling to the environment seems justified.



\end{document}